\documentclass[conference]{IEEEtran}
\usepackage{graphicx}  
\usepackage{amsmath,amssymb}  
\usepackage{cite}  
\usepackage{url}

\IEEEoverridecommandlockouts

\usepackage[utf8]{inputenc}
\usepackage{listings}
\usepackage{xcolor}
\usepackage{graphicx} 
\usepackage{amsmath} 

\newcommand{\zeroState}{|0\rangle}   
\newcommand{\oneState}{|1\rangle}    
\newcommand{\plusState}{|+\rangle}   
\newcommand{\minusState}{|-\rangle}  
\newcommand{\plusYState}{|\!+i\rangle} 
\newcommand{\minusYState}{|\!-i\rangle} 

\newcommand{\pzZero}{p_{\text{Z}}(+1)}   
\newcommand{\pzOne}{p_{\text{Z}}(-1)}    
\newcommand{\pxPlus}{p_{\text{X}}(+1)}

\def \btheta {\boldsymbol{\theta}}

\global\long\def\RR{\mathbb{R}}
\global\long\def\CC{\mathbb{C}}

\newcommand{\bfM}{\mathbf{M}}

\newcommand{\bfX}{\mathbf{X}}
\newcommand{\bfY}{\mathbf{Y}}
\newcommand{\bfZ}{\mathbf{Z}}

\def \tensor {\otimes}

\def\<{\langle}
\def\>{\rangle}

\usepackage{stackengine}
\def \deq {:=}

\def \eqand {\text{ and }}

\begin{document}

\title{QubitLens: An Interactive Learning Tool for Quantum State Tomography 
}

\author{
\IEEEauthorblockN{Mohammad Aamir Sohail }
\IEEEauthorblockA{University of Michigan, USA \\
Email: mdaamir@umich.edu  
}
\and
\IEEEauthorblockN{Ranga Sudharshan}
\IEEEauthorblockA{ University of Michigan, USA \\
Email: rsudhars@umich.edu 
}
\and
\IEEEauthorblockN{S. Sandeep Pradhan}
\IEEEauthorblockA{University of Michigan, USA \\
Email: pradhanv@umich.edu
}
\and
\IEEEauthorblockN{Arvind Rao}
\IEEEauthorblockA{University of Michigan, USA \\
Email:ukarvind@umich.edu  
}
}

\maketitle

\begin{abstract}

Quantum state tomography is a fundamental task in quantum computing, involving the reconstruction of an unknown quantum state from measurement outcomes. Although essential, it is typically introduced at the graduate level due to its reliance on advanced concepts such as the density matrix formalism, tensor product structures, and partial trace operations. This complexity often creates a barrier for students and early learners. In this work, we introduce QubitLens, an interactive visualization tool designed to make quantum state tomography more accessible and intuitive. QubitLens leverages maximum likelihood estimation (MLE), a classical statistical method, to estimate pure quantum states from projective measurement outcomes in the $\bfX$, $\bfY$, and $\bfZ$ bases. The tool emphasizes conceptual clarity through visual representations, including Bloch sphere plots of true and reconstructed qubit states, bar charts comparing parameter estimates, and fidelity gauges that quantify reconstruction accuracy. QubitLens offers a hands-on approach to learning quantum tomography without requiring deep prior knowledge of density matrices or optimization theory. The tool supports both single- and multi-qubit systems and is intended to bridge the gap between theory and practice in quantum computing education.
\end{abstract}



\section{Introduction}

Quantum computing is an emerging technology with the potential to tackle problems that are intractable for classical systems. A fundamental requirement in quantum computing is the ability to accurately prepare, manipulate, and measure quantum states. One critical task in this context is \emph{quantum state tomography}, a foundational concept in quantum information theory.  It involves reconstructing an unknown quantum state by analyzing the outcomes from a set of measurements performed on multiple identical copies of the state.

Quantum state tomography is most commonly introduced in graduate-level courses, as it involves several technically demanding concepts. These include the density matrix formalism, partial trace operations, quantum channels, and the generalized framework of quantum measurement ~\cite{nielsen_chuang_2010, paris2004quantum, wilde2013quantum}. While foundational, these concepts can pose a steep learning curve for students who are just beginning to learn about quantum computing.
There have been numerous well-established methods for quantum state tomography, including linear inversion~\cite{dariano2003quantum}, Bayesian estimation~\cite{blume2010optimal}, and maximum likelihood estimation (MLE)~\cite{hradil1997quantum, rehacek2007diluted}. Other approaches, such as compressed sensing \cite{gross2010quantum} and neural-network-based tomography~\cite{torlai2017many}, have also been proposed to improve scalability and accuracy in high-dimensional settings. Although powerful, these methods often require an advanced background in quantum mechanics, statistics, and optimization theory, which makes them less approachable for beginners or for educational purposes.

In this work, we introduce QubitLens, an interactive and educational tool designed to help students and educators better understand the basic principles of quantum state tomography. QubitLens focuses on pure-state tomography for single and multi-qubit systems, using MLE as a backend algorithm. Importantly, the tool avoids complicated density matrix manipulations and instead emphasizes intuitive, visual elements such as Bloch sphere representations and bar plots to assess reconstruction quality. By making the learning process hands-on and visually guided, QubitLens offers a more approachable path for beginners to engage with one of the most essential ideas in quantum information theory.

This paper is structured as follows. We first introduce the concept of MLE in a simple and accessible way. Since MLE is a common statistical method often taught in high school or early undergraduate courses, we show how it can be naturally extended to quantum state estimation. Next, we provide the necessary background in quantum computing, with a particular focus on measurements performed in different Pauli bases for a single qubit. Following this, we present QubitLens, and finally, we extend our framework to accommodate the multi-qubit scenario.

\section{Overview of Maximum Likelihood Estimation}\label{sec:MLE}

Maximum likelihood estimation (MLE) is a fundamental statistical technique used to estimate unknown parameters of a probability distribution from observed data. Consider a parametric family of probability distributions $\{P_{\boldsymbol{\theta}}\}$ indexed by a parameter vector $\boldsymbol{\theta}$. We have given a dataset $\mathcal{D}$ which is assumed to be independently and identically distributed (i.i.d.) according to $P_{\boldsymbol{\theta}_{\text{true}}}$, where $\boldsymbol{\theta}_{\text{true}}$ is the true but unknown parameter. The goal of MLE is to find the parameter $\hat{\boldsymbol{\theta}}$ that maximizes the likelihood of the observed data. Formally,
\[
\hat{\boldsymbol{\theta}} = \arg\max_{\boldsymbol{\theta}} \mathcal{L}(\boldsymbol{\theta}; \mathcal{D}),
\]
where $\mathcal{L}(\boldsymbol{\theta}; \mathcal{D})$ denotes the likelihood function, i.e., the probability of observing the data $\mathcal{D}$ under the distribution parameterized by $\boldsymbol{\theta}$. In practice, the logarithm of the likelihood, known as the log-likelihood, is typically used instead of the likelihood for both analytical and numerical convenience.

To illustrate this process concretely, we consider three mutually independent binary random variables $\mathsf{X}$, $\mathsf{Y}$, and $\mathsf{Z}$, each taking values in $\{+1, -1\}$. We model these using Bernoulli random variables, where a success corresponds to observing $+1$. The probability of success for each variable is assumed to depend on a shared underlying parameter vector $\boldsymbol{\theta}$, with $p_{\text{X}}(\boldsymbol{\theta})$, $p_{\text{Y}}(\boldsymbol{\theta})$, and $p_{\text{Z}}(\boldsymbol{\theta})$ denoting the corresponding success probabilities.

Suppose we observe $N_{\text{X}}$, $N_{\text{Y}}$, and $N_{\text{Z}}$ i.i.d. samples according to $p_{\text{X}}(\boldsymbol{\theta})$, $p_{\text{Y}}(\boldsymbol{\theta})$, and $p_{\text{Z}}(\boldsymbol{\theta})$, respectively, denoted by:
\[
\mathsf{x}_1, \dots, \mathsf{x}_{N_{\text{X}}}; \quad \mathsf{y}_1, \dots, \mathsf{y}_{N_{\text{Y}}}; \quad \mathsf{z}_1, \dots, \mathsf{z}_{N_{\text{Z}}}.
\]
The number of observed successes for each variable is:
\[
n_{\text{X}} = \sum_{i=1}^{N_{\text{X}}}  \mathbf{1}_{\{\mathsf{x}_i = +1\}}, \
n_{\text{Y}} = \sum_{i=1}^{N_{\text{Y}}}  \mathbf{1}_{\{\mathsf{y}_i = +1\}}, \
n_{\text{Z}} = \sum_{i=1}^{N_{\text{Z}}}  \mathbf{1}_{\{\mathsf{z}_i = +1\}},
\]
where $\mathbf{1}_{\{\mathcal{E}\}}$ denotes the indicator function, which equals $1$ if the event $\mathcal{E}$ occurs and $0$ otherwise.
The likelihood function for each variable is based on the Bernoulli model:
\begin{align}
    \mathcal{L}_X(\boldsymbol{\theta};\mathsf{x}_i) &= p_{\text{X}}(\boldsymbol{\theta})^{n_{\text{X}}} (1 - p_{\text{X}}(\boldsymbol{\theta}))^{(N_{\text{X}} - n_{\text{X}})}, \label{eqn:Lx} \\
    \mathcal{L}_Y(\boldsymbol{\theta};\mathsf{y}_i) &= p_{\text{Y}}(\boldsymbol{\theta})^{n_{\text{Y}}} (1 - p_{\text{Y}}(\boldsymbol{\theta}))^{(N_{\text{Y}} - n_{\text{Y}})}, \label{eqn:Ly}\\
    \mathcal{L}_Z(\boldsymbol{\theta};\mathsf{z}_i) &= p_{\text{Z}}(\boldsymbol{\theta})^{n_{\text{Z}}} \ (1 - p_{\text{Z}}(\boldsymbol{\theta}))^{(N_{\text{Z}} - n_{\text{Z}})}. \label{eqn:Lz}
\end{align}
Since $\mathsf{x}_i$, $\mathsf{y}_i$, and $\mathsf{z}_i$ are i.i.d., the joint likelihood over all observations is simply the product of the individual likelihoods:
\begin{equation}\label{eqn:likelihood}
    \mathcal{L}(\boldsymbol{\theta}; \mathsf{x}_i,\mathsf{y}_i,\mathsf{z}_i) = \mathcal{L}_X(\boldsymbol{\theta};\mathsf{x}_i) \cdot \mathcal{L}_Y(\boldsymbol{\theta};\mathsf{y}_i) \cdot \mathcal{L}_Z(\boldsymbol{\theta};\mathsf{z}_i).
\end{equation}

Taking the logarithm yields the total log-likelihood:
\begin{align*}
\log \mathcal{L}(\boldsymbol{\theta}) = \ 
&\ n_{\text{X}} \log p_{\text{X}}(\boldsymbol{\theta}) + (N_{\text{X}} - n_{\text{X}})\log(1 - p_{\text{X}}(\boldsymbol{\theta})) \\
+ &\ n_{\text{Y}} \log p_{\text{Y}}(\boldsymbol{\theta}) + (N_{\text{Y}} - n_{\text{Y}})\log(1 - p_{\text{Y}}(\boldsymbol{\theta})) \\
+ &\ n_{\text{Z}} \log p_{\text{Z}}(\boldsymbol{\theta}) \ + (N_{\text{Z}} - n_{\text{Z}})\log(1 - p_{\text{Z}}(\boldsymbol{\theta})).
\end{align*}

To estimate the parameter vector $\boldsymbol{\theta}$, one maximizes the log-likelihood function using numerical optimization methods. Depending on the form of $p_{\text{X}}(\boldsymbol{\theta})$, $p_{\text{Y}}(\boldsymbol{\theta})$, and $p_{\text{Z}}(\boldsymbol{\theta})$, this optimization may be performed using gradient-based methods (e.g., gradient descent or ADAM) or gradient-free methods such as COBYLA or L-BFGS \cite{nocedal1999numerical}. These algorithms iteratively update the $\btheta$ (parameter estimate) to approach the $\btheta_{\text{true}}$.

\section{Fundamental Idea of Qubit}
A qubit is the quantum counterpart of a classical bit, capable of existing in a coherent superposition of the binary values 
$0$ and 
$1$. Unlike a classical bit, which deterministically assumes a value of either $0$ or $1$ at any given time, a qubit is described by a column vector in a two-dimensional complex Hilbert space. Mathematically, a single qubit state is expressed as a linear combination of two orthonormal basis states, conventionally denoted in Dirac notation as $|0\>\eqand |1\>$, with corresponding column vector representations:
$$|0\> = \begin{bmatrix}
    1\\0
\end{bmatrix}\ \eqand |1\> = \begin{bmatrix}
    0\\1
\end{bmatrix}.$$
The orthonormal basis states $|0\>\eqand |1\>$ are commonly referred to as the computational basis. An arbitrary (pure) qubit state can be written as a superposition of these basis states:
$$|\psi\> = \alpha |0\> +\beta |1\> = \begin{bmatrix}
    \alpha\\\beta
\end{bmatrix}, $$
where $\alpha,\beta \in \CC$ are complex probability amplitudes satisfying the normalization condition:  $|\alpha|^2 + |\beta|^2 = 1$. The conjugate transpose of this state, denoted as $\langle \psi|$, is given as a row vector $$\langle \psi| := [\alpha^* \ \beta^*],$$ where $\alpha^*$ and $\beta^*$ are the complex conjugate of $\alpha \eqand \beta$, respectively. This row vector form allows one to compute expressions like the inner product 
$\langle\psi|\psi\rangle = |\alpha|^2+|\beta|^2 = 1$, which confirms that the state is properly normalized.
\subsection{Quantum Measurements} In quantum mechanics, we can not directly ``see"  the value of $\alpha$ and $\beta$. Instead, to extract information about the qubit, we must \textit{measure} it. 
A quantum measurement is described by a set of projection operators $\{P_i\}$, which satisfy the following properties: $1.$ (Hermitian) $P_i = P_i^\dagger$, $2.$ (Idempotent) $P_i^2 = P_i$, and $3.$ (Completeness) $\sum_i P_i = I$.

For a single qubit measured in the computational basis, the measurement answers the question: "Is the qubit in state $|0\rangle$ or $|1\rangle$?" We can describe the measurement in computational basis in terms of the Pauli-$\bfZ$ observable: $$\bfZ = |0\rangle\langle 0| - |1\rangle\langle 1|,$$ which has eigenvalues $+1$ and $-1$, corresponding to the computational basis states $|0\rangle$ and $|1\rangle$, respectively. The associated projectors onto the $\pm 1$ eigenspaces are $$P_{+1} = |0\rangle\langle 0|\eqand  P_{-1} = |1\rangle\langle 1|.$$ Thus, the probabilities of observing eigenvalue $+1$ or $-1$ when measuring in $\bfZ$ (or computational) basis are \begin{align*}
    \text{Prob}(+1) &= \langle \psi | P_{+1} | \psi \rangle = |\<0|\psi\>|^2 =  |\alpha|^2,\\
     \text{Prob}(-1) &= \langle \psi | P_{-1} | \psi \rangle = |\<1|\psi\>|^2 =  |\beta|^2.
\end{align*}
Immediately after the measurement, the quantum state collapses to the observed outcome. For example, if the result is 
$+1$, the state $|\psi\>$ is then collapsed to state
$|0\>$; the original superposition is irreversibly lost.

Following the above discussion, it is clear that a measurement process is fundamentally probabilistic, which means that a single observation cannot reveal the full structure of the quantum state. In practice, the quantum measurement is often executed repeatedly on identically prepared qubits (a process referred to as taking multiple shots) to collect a distribution of measurement outcomes. This empirical distribution provides estimates of the underlying probabilities 
$|\alpha|^2$
  and 
$|\beta|^2$, which can be used for reconstructing the quantum state (or quantum state tomography).
\subsection{Bloch Sphere Representation of Qubits}
\begin{figure}[ht]
    \centering
    \includegraphics[scale=0.65]{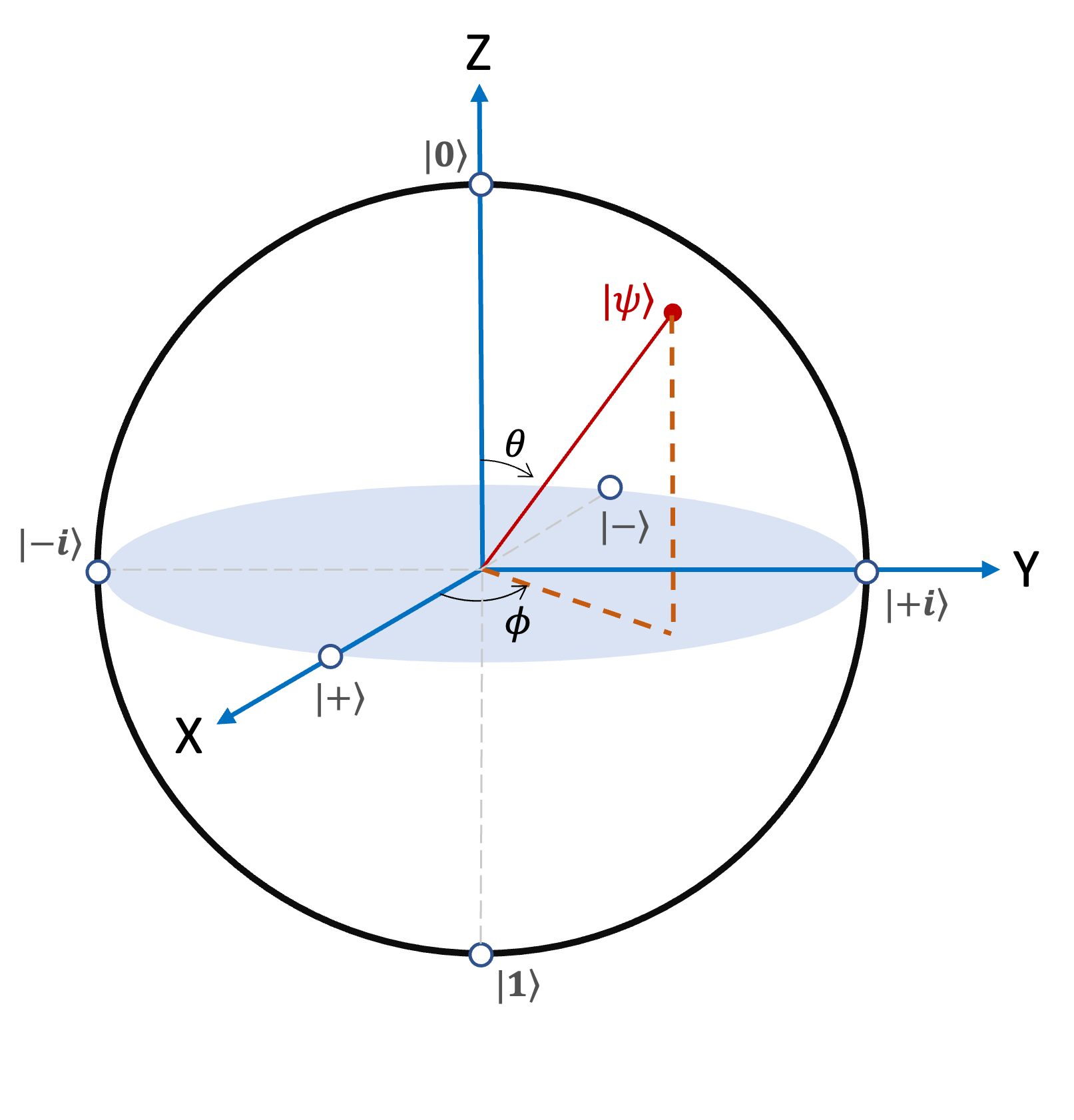}
    \caption{Bloch sphere}
    \label{fig:bloch_eg}
\end{figure}

The Bloch sphere is a powerful and widely used geometric tool for visualizing the state of a single qubit. It provides an intuitive geometric representation in which any qubit state can be uniquely mapped to a point on the surface of a unit sphere in $\RR^3$. At first, it seems that we require four real parameters (two for each complex probability amplitude) to describe a qubit. However, taking out the common phase factor from $\alpha$ and $\beta$ (referred to as global phase\footnote{The global phase has no meaningful physical significance and can be ignored.}) and the normalization condition, we need only two real parameters to fully characterize a qubit's physical state. In the Bloch sphere representation, the polar angle $\theta$ and the azimuthal angle $\phi$ serve as two real parameters that fully characterize the state of a qubit. The probability amplitudes are expressed in terms of $\theta$ and $\phi$ as:
$\alpha = \cos\left({\theta}/{2}\right)\eqand \beta = e^{i\phi}\sin\left({\theta}/{2}\right),$
where $\theta \in [0,\pi]$ and $\phi \in [0,2\pi)$. It can be easily verified that the parameterization satisfies the condition $|\alpha|^2 + |\beta|^2 = 1$. Therefore, the state $|\psi\rangle$ can be rewritten as:
$$|\psi\> =\cos\left(\frac{\theta}{2}\right) |0\> + e^{i\phi}\sin\left(\frac{\theta}{2}\right) |1\>.$$ This parameterization aligns with spherical coordinates and uniquely locates the qubit state on the surface of the Bloch sphere. To relate this to $\RR^3$, we can translate the angles into Cartesian coordinates:
$x= \sin(\theta)\cos(\phi), y=\sin(\theta)\sin(\phi), \eqand z = \cos(\theta)$.
This 3D point 
$(x,y,z)$ provides a complete and visual description of the qubit state (see Fig.~\ref{fig:bloch_eg}). The vector from the origin to this 3D point is referred to as the Bloch vector.
\subsection{Measurement in Pauli Basis}\label{subsec:meas_Pauli}
The outcome of a quantum measurement depends on both the qubit state and the measurement basis. A qubit state, represented by a point on the Bloch sphere, is projected onto a specific axis determined by the eigenbasis of the observable being measured. The measurement collapses the qubit onto one of the eigenbases with a probability dictated by the overlap between the state and the eigenbasis. 
For example, measuring in the Pauli-\( \mathbf{Z} \), \( \mathbf{X} \), or \( \mathbf{Y} \) basis projects a state onto the eigenbasis
\(\{|0\rangle, |1\rangle\}\), 
\(\{|\pm\rangle \deq  (|0\rangle \pm |1\rangle)/\sqrt{2}\}\), or 
\(\{|\pm i\rangle \deq (|0\rangle \pm i|1\rangle)/\sqrt{2}\}\), respectively. These eigenbases align with the \( z \)-, \( x \)-, and \( y \)-axes of the Bloch sphere, respectively. In each case, the measurement outcome is either \( +1 \) or \( -1 \), corresponding to the first and second states in the respective basis.


\subsubsection*{Measurement in the $\bfZ$-basis }
When measuring in $\bfZ$-basis, the state $|\psi\>$ collapse to either \(\zeroState\) or \(\oneState\) with the probability given as,
$\pzZero \deq |\<0|\psi\>|^2$ and $\pzOne \deq |\<1|\psi\>|^2$, respectively. After simplifying the expression and using the trigonometric identities, we get probabilities in terms of Bloch sphere parameters $\theta$ and $\phi$ 
\begin{align*}
    &\pzZero = |\cos(\theta/2)|^2 = (1+\cos(\theta))/2 \eqand\\
\pzOne &= |e^{i\phi}\sin(\theta/2)|^2 =\sin^2(\theta/2) = (1-\cos(\theta))/2.
\end{align*}
\subsubsection*{Measurement in the $\bfX$-basis}
When measuring in $\bfX$-basis, the state $|\psi\rangle$ collapses to either \(\plusState\) or \(\minusState\) with probabilities defined as $p_{\text{X}}(+1) \deq |\<+|\psi\>|^2$ and $p_{\text{X}}(-1) \deq |\<-|\psi\>|^2$, respectively. We can further simplify these expressions.
\begin{align*}
    \pxPlus  
    &= |\cos(\theta/2)\<+|0\> + e^{i\phi}\sin(\theta/2) \<+|1\>|^2\\
    &\overset{a}{=} \frac{1}{{2}}|\cos(\theta/2) + e^{i\phi}\sin(\theta/2)|^2\\
    &\overset{}{=} \frac{1}{{2}}(\cos^2(\theta/2) + \sin^2(\theta/2) \\
    & \hspace{50pt} + (e^{i\phi}+e^{-i\phi})\sin(\theta/2)\cos(\theta/2))\\
    &\overset{b}{=} \frac{1}{{2}}(1 + \cos(\phi)\sin(\theta))
\end{align*}
where $(a)$ follows because $\<+|0\> = \<+|1\> = 1/\sqrt{2}$ and $(b)$ follows from  trigonometric identities.
Similarly, $p_{\text{X}}(-1)$ can also be derived. However, using the fact that $p_{\text{X}}(+1) + p_{\text{X}}(-1) = 1$, we can easily get $p_{\text{X}}(-1) = (1 - \cos(\phi)\sin(\theta))/2$.







\subsubsection*{Measurement in the $\bfY$-basis} 
When measuring in $\bfY$-basis, the state $|\psi\rangle$ collapses to either \(\plusYState\) or \(\minusYState\) with probabilities defined as $p_{\text{Y}}(+1) \deq |\<+i|\psi\>|^2$ and $p_{\text{Y}}(-1) \deq |\<-i|\psi\>|^2$, respectively. We can further simplify these expressions.
\begin{align*}
   p_{\text{Y}}(+1)  
    &= |\cos(\theta/2)\<+i|0\> + e^{i\phi}\sin(\theta/2) \<+i|1\>|^2\\
    &\overset{}{=} \frac{1}{{2}}|\cos(\theta/2) - i e^{i\phi}\sin(\theta/2)|^2\\
    &\overset{}{=} \frac{1}{{2}}(\cos^2(\theta/2) + \sin^2(\theta/2) \\
    & \hspace{50pt} + i(e^{-i\phi}-e^{i\phi})\sin(\theta/2)\cos(\theta/2))\\
    &\overset{}{=} \frac{1}{{2}}(1 + \sin(\phi)\sin(\theta))
\end{align*}
Using the fact that $p_{\text{Y}}(+1) + p_{\text{Y}}(-1) = 1$, we get $p_{\text{Y}}(-1) = (1 - \sin(\phi)\sin(\theta))/2$.

Although each Pauli basis measurement yields only a binary outcome, the choice of basis determines which component of the qubit's Bloch vector is probed. Measuring in the $\bfZ$-basis provides access to the polar angle \( \theta \), which describes how close the state lies to the \( |0\rangle \) or \( |1\rangle \) poles. However, this measurement gives no information about the azimuthal angle \( \phi \), which determines the qubit's orientation in the x–y plane. To obtain partial information about \( \phi \), measurement in the $\bfX$-basis is used, which is sensitive to \( \cos\phi \). Yet, since \( \cos\phi = \cos(2\pi-\phi) \), the value of \( \phi \) remains ambiguous: two distinct points on the Bloch sphere yield the same measurement statistics in the $\bfX$-basis. This ambiguity is resolved by including $\bfY$-basis measurement, which is sensitive to \( \sin\phi \) and thus breaks the symmetry, allowing us to determine the exact azimuthal angle.

Together, measuring multiple identical copies of a qubit in the $\bfX$, $\bfY$, and $\bfZ$ bases provides complete information about the Bloch sphere coordinates \( (\theta, \phi) \) of a pure qubit state. This forms the foundation of \emph{quantum state tomography}. In the following section, we describe how single-qubit state tomography is performed using this observation and MLE.

\section{Single-Qubit State Tomography using MLE}\label{sec:STusingMLE}

We now present the algorithm for performing single-qubit state tomography using the MLE. Alongside the conceptual description, we provide corresponding code snippets for each step to enhance understanding. The code is implemented in Python 3 using \texttt{NumPy}, \texttt{matplotlib}, \texttt{ipywidgets}, and \texttt{QuTiP} libraries. 

The algorithm takes the true parameters $\theta_{\text{true}}$ and $\phi_{\text{true}}$ as inputs, along with the number of measurement shots $N_{\text{X}},N_{\text{Y}}, \eqand N_{\text{Z}}$ for the $\bfX,\bfY,\eqand \bfZ$ bases, respectively. Below, we detail each step involved in the algorithm.

\vspace{5pt}
   \noindent \textbf{1. Qubit State Preparation}:
    Given the true parameters $(\theta_{\text{true}}, \phi_{\text{true}})$, we prepare the following (pure) qubit state
    \[
    |\psi_{\text{true}}\rangle = \cos\left(\frac{\theta_{\text{true}}}{2}\right)|0\rangle + e^{i\phi_{\text{true}}}\sin\left(\frac{\theta_{\text{true}}}{2}\right)|1\rangle.
    \]
    The corresponding code to generate this state is:
    \begin{lstlisting}
import numpy as np

def create_state(theta, phi):
    return np.array([np.cos(theta/2), np.exp(1j * phi) * np.sin(theta/2)])

true_state = create_state(theta_true,phi_true)
    \end{lstlisting}

\vspace{5pt}
   \noindent \textbf{2. Measurement Simulation}:
    To simulate quantum measurements, we need to compute the probabilities of obtaining the $+1$ eigenvalue outcomes in the $\bfX,\bfY,\eqand \bfZ$ bases for $|\psi_{\text{true}}\>$, as discussed in Section \ref{subsec:meas_Pauli}. The probabilities are given as
    \[
    p(+1 \, | \, \text{basis}) = 
    \begin{cases}
    {(1 + \cos(\theta_{\text{true}}))}/{2}, & \text{for } \bfZ \text{ basis}, \\
    {(1 + \cos(\phi_{\text{true}}) \sin(\theta_{\text{true}}))}/{2}, & \text{for } \bfX \text{ basis}, \\
    {(1 + \sin(\phi_{\text{true}}) \sin(\theta_{\text{true}}))}/{2}, & \text{for } \bfY \text{ basis}.
    \end{cases}
    \]
    Once the probabilities are known, binary measurement outcomes can be simulated by randomly sampling $+1$ or $-1$ according to the computed probabilities.
The corresponding code to simulate measurement in Pauli bases is given below.
    \begin{lstlisting}
def meas_prob(state, basis):
    theta = np.real(2*np.arccos(state[0]))
    phi = np.angle(state[1])
    if basis == 'Z':
        return (1 + np.cos(theta)) / 2
    elif basis == 'X':
        return (1 + np.cos(phi) * np.sin(theta)) / 2
    elif basis == 'Y':
        return (1 + np.sin(phi) * np.sin(theta)) / 2
   
def simulate_meas(state, basis, n_shots):
    p1 = meas_prob(state, basis)
    p2 = 1 - p1
    return np.random.choice([+1, -1], size=n_shots, p=[p1, p2])

meas_x = simulate_meas(true_state, 'X', Nx)
meas_y = simulate_meas(true_state, 'Y', Nz)
meas_z = simulate_meas(true_state, 'Z', Nz)
    \end{lstlisting}
    

\vspace{5pt}
   \noindent  \textbf{3. Construct the Likelihood Function}:
    Let $\{\mathsf{x}_i\},\{\mathsf{y}_i\}$, and $\{\mathsf{z}_i\}$ be the  $N_{\text{X}}$, $N_{\text{Y}}$, and $N_{\text{Z}}$ measurement outcomes in the $\bfX,\bfY,\eqand\bfZ$ bases, respectively, generated from Step 2. 
The corresponding likelihood function can be written as (please refer to Section \ref{sec:MLE} for further details): 
    \begin{align*}
    \mathcal{L}_X(\boldsymbol{\theta};\mathsf{x}_i) &= p_{\text{X}}(\boldsymbol{\theta})^{n_{\text{X}}} (1 - p_{\text{X}}(\boldsymbol{\theta}))^{(N_{\text{X}} - n_{\text{X}})}, \\
    \mathcal{L}_Y(\boldsymbol{\theta};\mathsf{y}_i) &= p_{\text{Y}}(\boldsymbol{\theta})^{n_{\text{Y}}} (1 - p_{\text{Y}}(\boldsymbol{\theta}))^{(N_{\text{Y}} - n_{\text{Y}})}, \\
    \mathcal{L}_Z(\boldsymbol{\theta};\mathsf{z}_i) &= p_{\text{Z}}(\boldsymbol{\theta})^{n_{\text{Z}}} \ (1 - p_{\text{Z}}(\boldsymbol{\theta}))^{(N_{\text{Z}} - n_{\text{Z}})}. 
\end{align*}
    Since measurements across different bases are independent, the overall likelihood can be written as the product of the individual likelihood functions:
    \[
    \mathcal{L}(\boldsymbol{\theta}; \mathsf{x}_i,\mathsf{y}_i,\mathsf{z}_i) = \mathcal{L}_X(\boldsymbol{\theta};\mathsf{x}_i) \cdot \mathcal{L}_Y(\boldsymbol{\theta};\mathsf{y}_i) \cdot \mathcal{L}_Z(\boldsymbol{\theta};\mathsf{z}_i).
    \]
    In practice, we work with the log-likelihood to ensure numerical stability. The corresponding code is outlined below:
    \begin{lstlisting}
def log_likelihood(params, samples_x, samples_y, samples_z):
    theta, phi = params
    state = create_state(theta, phi)
    
    log_L = 0
    for basis, samples in zip(['X', 'Y', 'Z'], [samples_x, samples_y, samples_z]):
        p1 = meas_prob(state, basis)
        p2 = (1 - p1)
        n_out1 = np.sum(samples == +1)
        n_out2 = np.sum(samples == -1)
        log_L += 
        n_out1 * np.log(p1 + 1e-10) + 
        n_out2 * np.log(p2 + 1e-10)
    return log_L
    \end{lstlisting}

\vspace{5pt}
   \noindent \textbf{4. MLE Estimation}:
   To find the maximum likelihood estimates \((\hat{\theta}, \hat{\phi})\), we need to minimize the negative log-likelihood function, represented as \(-\log \mathcal{L}(\boldsymbol{\theta}; \mathsf{x}_i, \mathsf{y}_i, \mathsf{z}_i)\). This can be done using a numerical optimization algorithm, such as L-BFGS, while adhering to the following constraints on the parameter ranges: $0 \leq \theta \leq \pi \eqand 0 \leq \phi < 2\pi.$ Starting from an initial guess, the optimizer will iteratively update the values of \((\theta, \phi)\) to best fit the observed measurement data \((\texttt{meas\_x}, \texttt{meas\_y}, \texttt{meas\_z})\) obtained from Step 2.

\begin{lstlisting}
def MLE(samples_x, samples_y, samples_z):
    opt_val = []
    opt_params = []
    theta_init = np.random.uniform(0, np.pi)
    phi_init = np.random.uniform(0, 2*np.pi)
    initial_params = [theta_init, phi_init]
    result = minimize(-log_likelihood, initial_params, args=(samples_x, samples_y, samples_z), bounds=[(0, np.pi), (0, 2*np.pi)], method='L-BFGS-B')
    return result.x

est_theta,est_phi = MLE(meas_x,meas_y,meas_z)
est_state = create_state(est_theta, est_phi)
\end{lstlisting}

\vspace{5pt}
   \noindent \textbf{5. Evaluate the Reconstruction}:
Using the estimated parameters \((\hat{\theta}, \hat{\phi})\), we reconstruct the estimated state \(|\hat{\psi}\rangle\). To assess the quality of this reconstruction, we can calculate the fidelity between the true state and the estimated state, defined as:
\[
F = |\langle \psi_{\text{true}} | \hat{\psi} \rangle|^2,
\]
where \(F = 1\) indicates perfect reconstruction.

\section{Interactive Visualization via QubitLens}
\label{sec:qubitlens}

\begin{figure}[!htb]
    \centering
    \includegraphics[width=\linewidth]{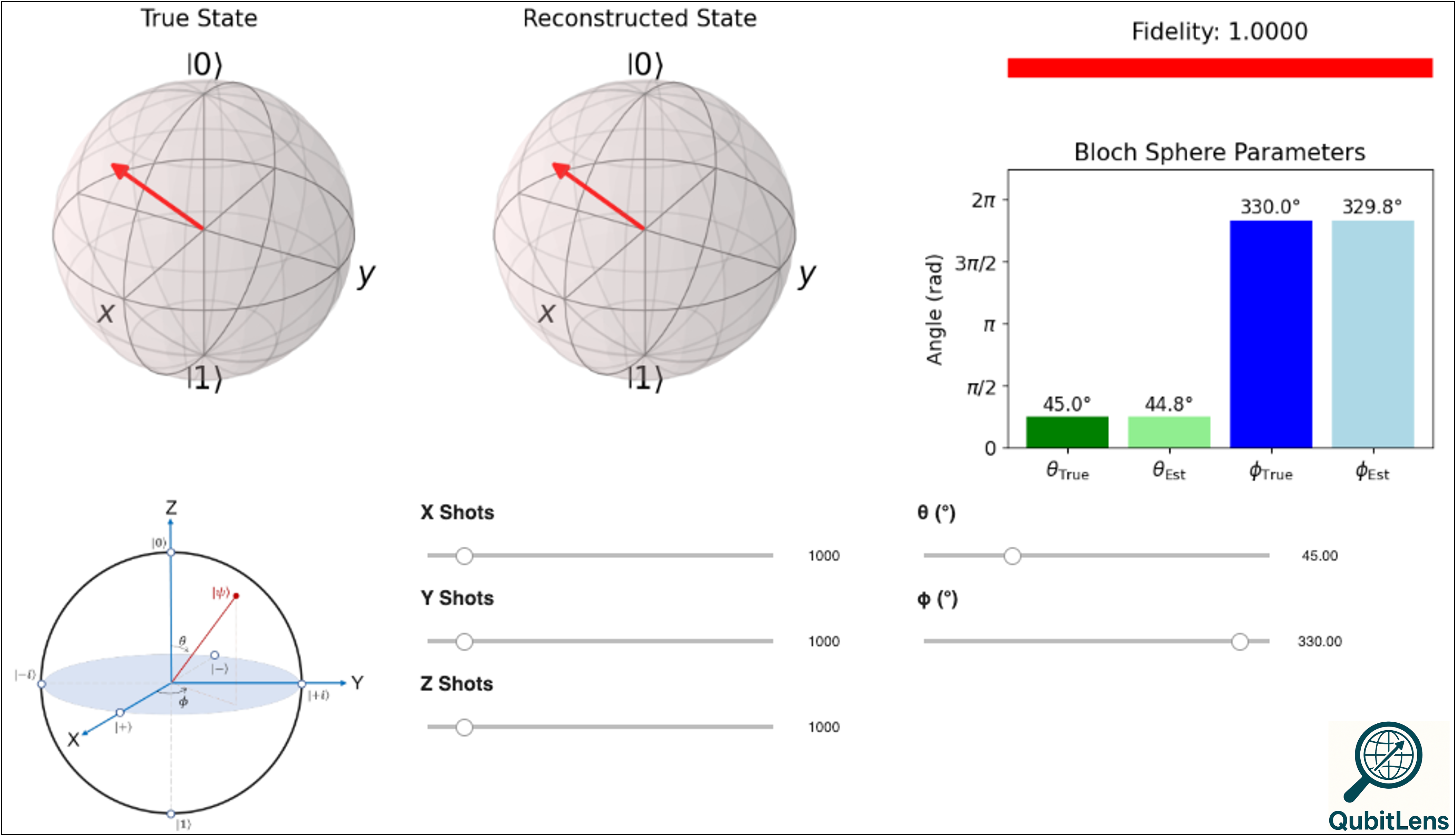}
    \caption{The {QubitLens} interface for single-qubit tomography. 
    The true and reconstructed states are shown on Bloch spheres (left), 
    with fidelity and parameter comparison (right). 
    Users can adjust the true parameters and the number of measurements in the \(\bfX\), \(\bfY\), and \(\bfZ\) bases using sliders, with all visualizations updating in real-time.}
    \label{fig:qubitLens}
\end{figure}

We embed the above qubit state tomography using the MLE algorithm within an interactive visualization tool we refer to as {QubitLens}. This tool is designed to provide intuitive insights into the process of single-qubit state tomography by exposing the algorithmic workflow through an interactive front end. Users can specify the true state parameters \((\theta_{\text{true}}, \phi_{\text{true}})\) using sliders, and independently choose the number of measurement shots in each Pauli basis (\(N_{\text{X}}, N_{\text{Y}}, N_{\text{Z}}\)). Upon adjusting these inputs, the interface dynamically updates three visual components: (i) a side-by-side Bloch sphere visualization of the true state and the reconstructed state, (ii) a bar plot comparing the true and estimated values of \(\theta\) and \(\phi\), and (iii) a fidelity gauge indicating the closeness of the reconstructed state to the true state (see Fig.~\ref{fig:qubitLens}).

\begin{figure}
    \centering
    \includegraphics[width=\linewidth]{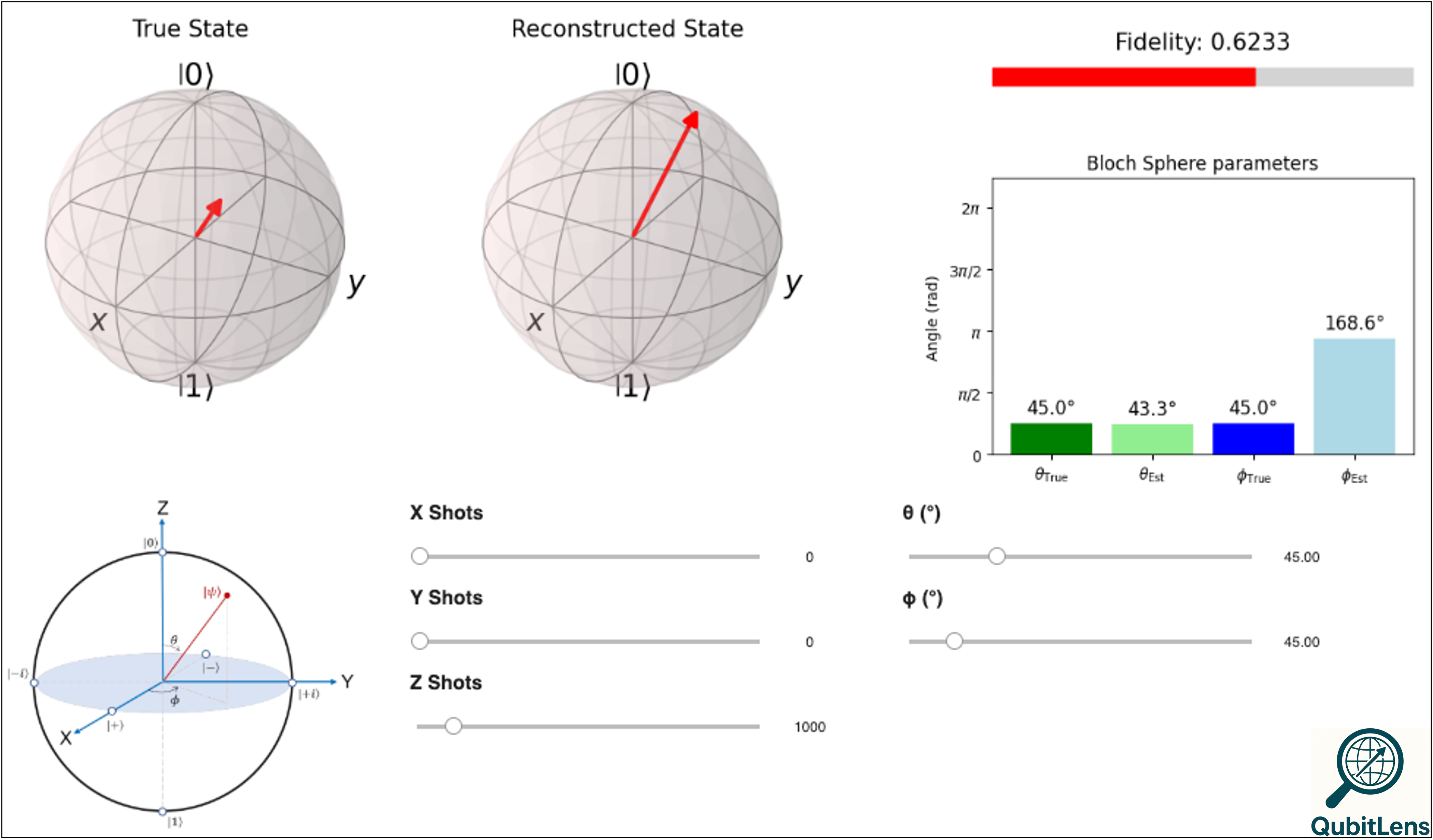}
    \caption{QubitLens output when measurements are restricted to the \(\bfZ\) basis. 
    Only the polar angle \(\theta\) is accurately estimated, while the azimuthal angle \(\phi\) remains unconstrained.}
    \label{fig:QubitLensZmeas}
\end{figure}

\begin{figure}
    \centering
    \includegraphics[width=\linewidth]{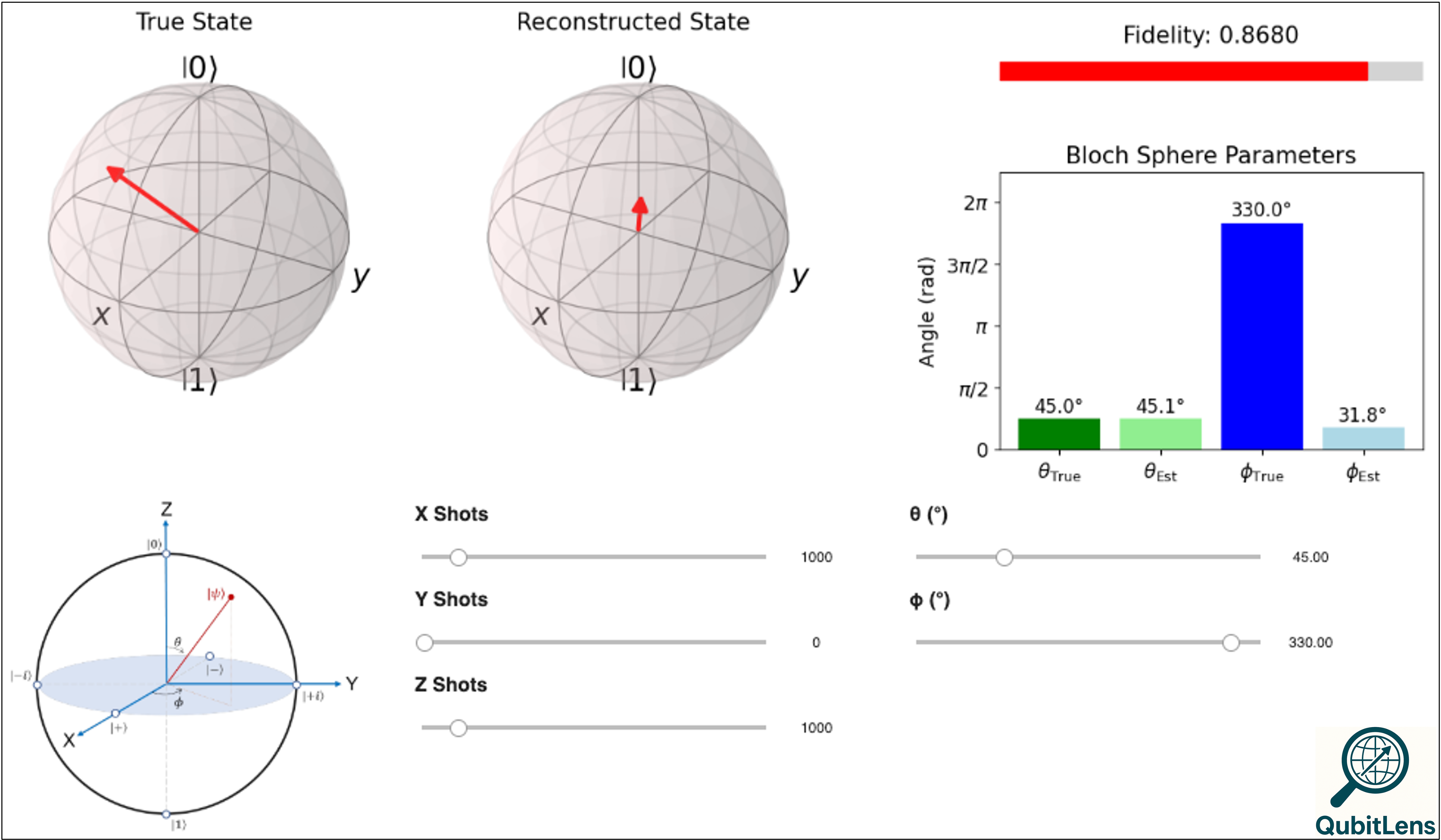}
\includegraphics[width=\linewidth]{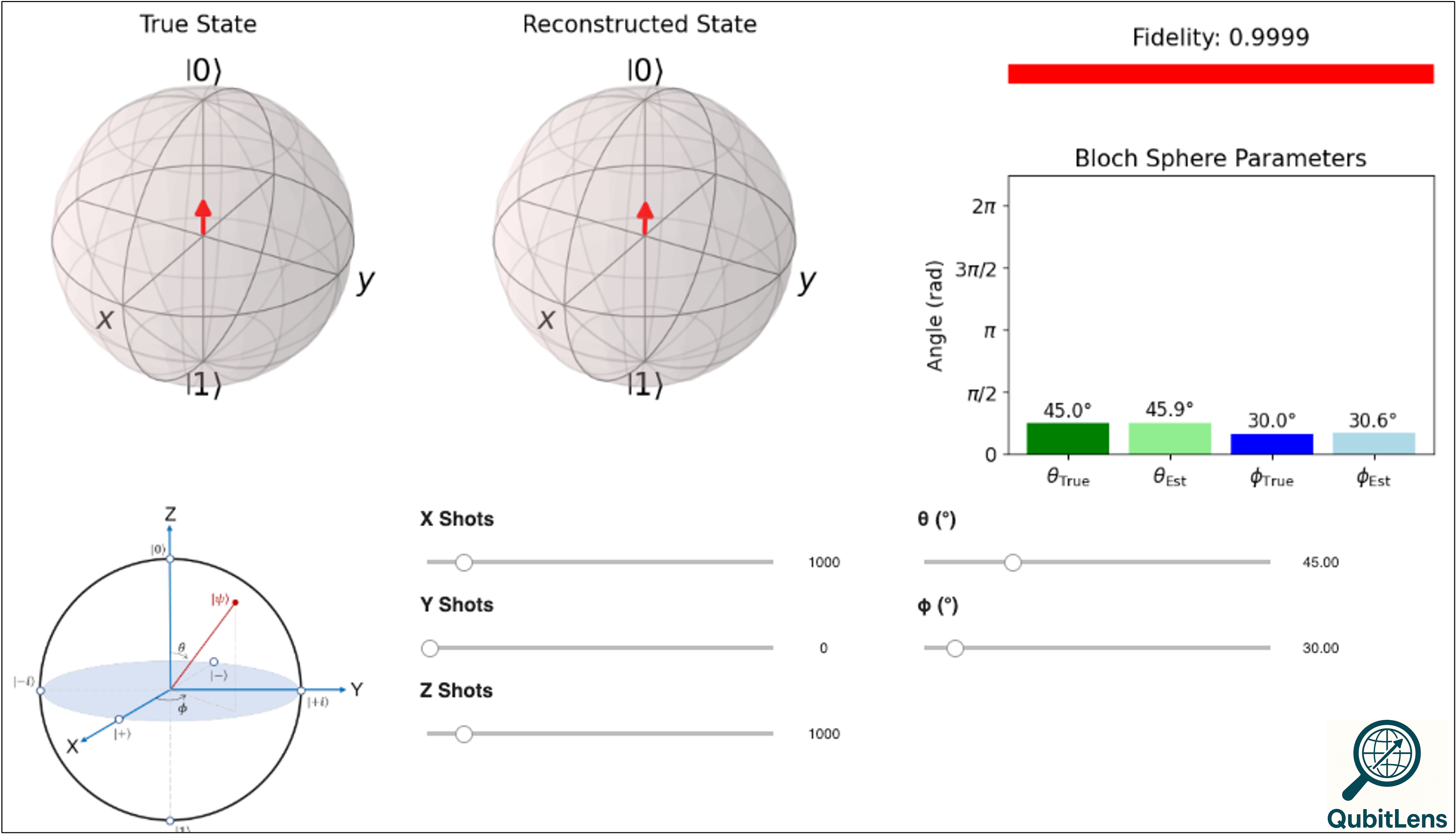}
\caption{Demonstration of ambiguity in estimating \(\phi\) when only \(\bfX\) and \(\bfZ\) basis measurements are used. 
    Top: the MLE outputs \(\hat{\phi}\! \approx \!(2\pi\! -\! \phi_{\text{true}}) \!= \!(2\pi\!-\!330^\circ) \!= \!30^{\circ}\). 
    Bottom: the MLE returns \(\hat{\phi} \approx \phi_{\text{true}}\!=\!30^{\circ}\). 
    }
    \label{fig:QubitLensXZmeas}
\end{figure}

The {Qubit Lens} framework allows users to explore how measurement choices affect the reconstruction of quantum states. For instance, as detailed in Section \ref{subsec:meas_Pauli}, when measurements are restricted to the \(\bfZ\) basis alone, only information about \(\theta\) can be recovered; the reconstructed state is necessarily insensitive to variations in \(\phi\) as shown in Fig.~\ref{fig:QubitLensZmeas}.
When measurements are taken in both the \(\bfX\) and \(\bfZ\) bases, the polar angle \(\theta\) can be reliably estimated, but ambiguity still remains in determining the azimuthal angle \(\phi\). In particular, the tomography algorithm can yield either \(\hat{\phi} = \phi_{\text{true}}\) or \(\hat{\phi} = 2\pi - \phi_{\text{true}}\). This behavior is illustrated in Fig.~\ref{fig:QubitLensXZmeas}, where the top part shows the case where the algorithm outputs \(\hat{\phi} = 2\pi - \phi_{\text{true}}\), while in the bottom part the algorithm returns \(\hat{\phi} = \phi_{\text{true}}\). To accurately estimate the parameters, it is essential to include measurements from all three Pauli bases because measurements in the \(\bfY\) basis provide information about \(\sin(\phi)\), which helps to break the ambiguity and allows for the precise estimation of \(\phi\), as shown in Fig.~\ref{fig:qubitLens}.

\section{Multi-Qubit State Tomography using MLE}

In this section, we extend the algorithm to multi-qubit systems. To illustrate this extension, we first describe how pure multi-qubit states can be parametrized using generalized polar coordinates, which form the foundation for the likelihood-based reconstruction procedure.

An arbitrary pure state of $n$ qubits can be represented as a unit vector in a $ 2^n-$dimensional complex Hilbert space. To ensure normalization and avoid redundancy due to the global phase, the state can be parametrized using a sequence of generalized polar coordinates as:
\begin{align*}
    &|\psi(\boldsymbol{\theta}, \boldsymbol{\phi})\rangle = 
\cos\left(\frac{\theta_1}{2}\right) |0\cdots00\rangle \\
&+ \sin\left(\frac{\theta_1}{2}\right)\cos\left(\frac{\theta_2}{2}\right) e^{i\phi_1}|0\cdots01\rangle
+ \cdots \\
&+ \sin\left(\frac{\theta_1}{2}\right)\cdots\sin\left(\frac{\theta_{(c-1)}}{2}\right)\cos\left(\frac{\theta_{c}}{2}\right)e^{i\phi_{(c-1)}}|1\cdots10\rangle\\
&\hspace{25pt}+ \sin\left(\frac{\theta_1}{2}\right)\sin\left(\frac{\theta_2}{2}\right)\cdots\sin\left(\frac{\theta_{c}}{2}\right)e^{i\phi_{c}}|1\cdots11\rangle,
\end{align*}
where $c = (2^n-1)$, the angles $\boldsymbol{\theta} = (\theta_1, \theta_2, \ldots, \theta_{c})$ and phases $\boldsymbol{\phi} = (\phi_1, \phi_2, \ldots, \phi_{c})$ completely specify the pure state, with $\theta_i \in [0, \pi]$ and $\phi_i \in [0, 2\pi)$. This parametrization ensures that the unit norm constraint \(\langle \psi (\boldsymbol{\theta},\boldsymbol{\phi}) | \psi (\boldsymbol{\theta},\boldsymbol{\phi}) \rangle = 1\) is automatically satisfied.

\subsection{Measurement Setting}

In single-qubit tomography, we perform measurements in the \(\bfX\), \(\bfY\), and \(\bfZ\) bases. In the multi-qubit setting, to fully reconstruct an \(n\)-qubit pure quantum state, we require measurement outcomes corresponding to all tensor products of Pauli operators (and identity operator), also referred to as \textit{Pauli strings}, on the \(n\)-qubit system.
There are \(4^n\) such Pauli strings, representing all combinations of \(\{\mathbf{I}, \bfX, \bfY, \bfZ\}^{\otimes n}\). Note that the all-identity string \(\mathbf{I}^{\otimes n}\) always yields a deterministic outcome of \(+1\) for pure states and thus provides no useful information. Therefore, we need to perform measurements corresponding to the remaining \(4^n - 1\) nontrivial Pauli strings. As an example, for \(n = 2\) qubits, there are \(4^2\!-\!1 = 15\) Pauli strings: $\mathbf{I}_1 \bfX_2, \mathbf{I}_1 \bfY_2, \mathbf{I}_1 \bfZ_2, \bfX_1 \mathbf{I}_2, \bfX_1 \bfX_2, \bfX_1 \bfY_2, \bfX_1 \bfZ_2,
\bfY_1 \mathbf{I}_2, \bfY_1 \bfX_2$,
$\bfY_1 \bfY_2, \bfY_1 \bfZ_2, \bfZ_1 \mathbf{I}_2, \bfZ_1 \bfX_2, \bfZ_1 \bfY_2,\eqand \bfZ_1 \bfZ_2.$
Here, \(A_1B_2\) is shorthand for the tensor product \(A \otimes B\), where the first qubit is measured in the \(A\) basis and the second in the \(B\) basis.

\subsection{Measurement Simulation}

Given a pure state $|\psi(\boldsymbol{\theta}, \boldsymbol{\phi})\rangle$ and a measurement setting $\bfM$ (a Pauli string), the probability of observing eigenvalue $+1$ is given by Born's rule:
\[
p(+1 \,|\, \bfM) = \langle \psi(\boldsymbol{\theta}, \boldsymbol{\phi}) | \Pi_{+}^\bfM | \psi(\boldsymbol{\theta}, \boldsymbol{\phi}) \rangle,
\]
where $\Pi_{+}^\bfM = \frac{1}{2}(\mathbf{I}^{\tensor n} + \bfM)$ is the projector onto the $+1$ eigenspace of $\bfM$. Similarly, the probability of observing eigenvalue $-1$ is:
\[
p(-1 \,|\, \bfM) = \langle \psi(\boldsymbol{\theta}, \boldsymbol{\phi}) | \Pi_{-}^\bfM | \psi(\boldsymbol{\theta}, \boldsymbol{\phi}) \rangle,
\]
where $\Pi_{-}^\bfM = \frac{1}{2}(\mathbf{I}^{\tensor n} - \bfM)$. Since $\Pi_{+}^\bfM + \Pi_{-}^\bfM = \mathbf{I}^{\tensor n}$, we have:
\[
p(-1 \,|\, \bfM) = 1 - p(+1 \,|\, \bfM).
\]
For each measurement basis $\bfM$, we collect $N_\text{M}$ independent binary measurement outcomes from the corresponding probability distribution.

\subsection{Construction of the Likelihood Function}

Suppose we perform $N_\text{M}$ measurement shots for each nontrivial Pauli strings $ \bfM$ and observe outcomes $\{m_1, m_2, \ldots, m_{N_\text{M}}\}$ where each $m_i \in \{+1, -1\}$. The likelihood of observing these outcomes under parameters $(\boldsymbol{\theta}, \boldsymbol{\phi})$ is:
\[
\mathcal{L}_M(\boldsymbol{\theta}, \boldsymbol{\phi}; \{m_i\}) = \prod_{i=1}^{N_\text{M}} p(m_i \,|\,  \bfM),
\]
where
\[
p(m_i \,|\, \bfM) =
\begin{cases}
p(+1 \,|\,  \bfM), & \text{if } m_i = +1, \\
p(-1 \,|\,  \bfM), & \text{if } m_i = -1.
\end{cases}
\]
Since all the measurements across different settings are independent, the overall likelihood can be written as the product of individual likelihoods:
\[
\mathcal{L}(\boldsymbol{\theta}, \boldsymbol{\phi}) = \prod_{\bfM \subseteq \{\mathbf{I}, \bfX, \bfY, \bfZ\}^{\otimes n} \backslash \mathbf{I}^{\otimes n} }\mathcal{L}_M(\boldsymbol{\theta}, \boldsymbol{\phi}; \{m_i\}),
\]
where the product is taken over all $(4^n\!-\!1)$ nontrivial Pauli operators.
For numerical stability, we maximize the log-likelihood written as:
\[
\log \mathcal{L}(\boldsymbol{\theta}, \boldsymbol{\phi}) = \sum_{\bfM\subseteq \{\mathbf{I}, \bfX, \bfY, \bfZ\}^{\otimes n} \backslash \mathbf{I}^{\otimes n}} \sum_{i=1}^{N_\text{M}} \log p(m_i \,|\, \bfM).
\]
\subsection{Interactive Visualization using QubitLens}

In the multi-qubit setting, the \textit{QubitLens} interface provides an interactive framework for visualizing the results of quantum state tomography at the level of individual qubits.
For multi-qubit systems, the pure state \( \in \mathbb{C}^{2^n}\) resides in a high-dimensional Hilbert space. To visualize individual qubits, we extract their reduced states via the partial trace operation \cite{nielsen_chuang_2010}. For the \(k\)th qubit, we compute the reduced density matrix \(\rho_k = \mathrm{Tr}_{\setminus k}(|\psi\rangle\langle\psi|)\), where the partial trace is taken over all qubits except the \(k\)th. This results in a single-qubit density matrix \(\rho_k \in \mathbb{C}^{2 \times 2}\), which can be uniquely represented by a three-dimensional real vector known as the Bloch vector.

The Bloch vector \(\vec{r}_k = (x_k, y_k, z_k)\) is defined by its projections onto the Pauli axes:
\[
x_k = \mathrm{Tr}(\rho_k \bfX), \quad
y_k = \mathrm{Tr}(\rho_k \bfY), \quad
z_k = \mathrm{Tr}(\rho_k \bfZ),
\]
where \(\bfX\), \(\bfY\), and \(\bfZ\) are the Pauli matrices. The resulting vector \(\vec{r}_k\) lies inside the unit ball in \(\mathbb{R}^3\), and it provides a geometric visualization of the qubit’s state. In particular, pure states correspond to Bloch vectors of unit length that lie on the surface of the sphere, while reduced states (or density matrix) have Bloch vectors strictly inside the sphere.

\begin{figure}[!h]
    \centering
    \includegraphics[width=\linewidth]{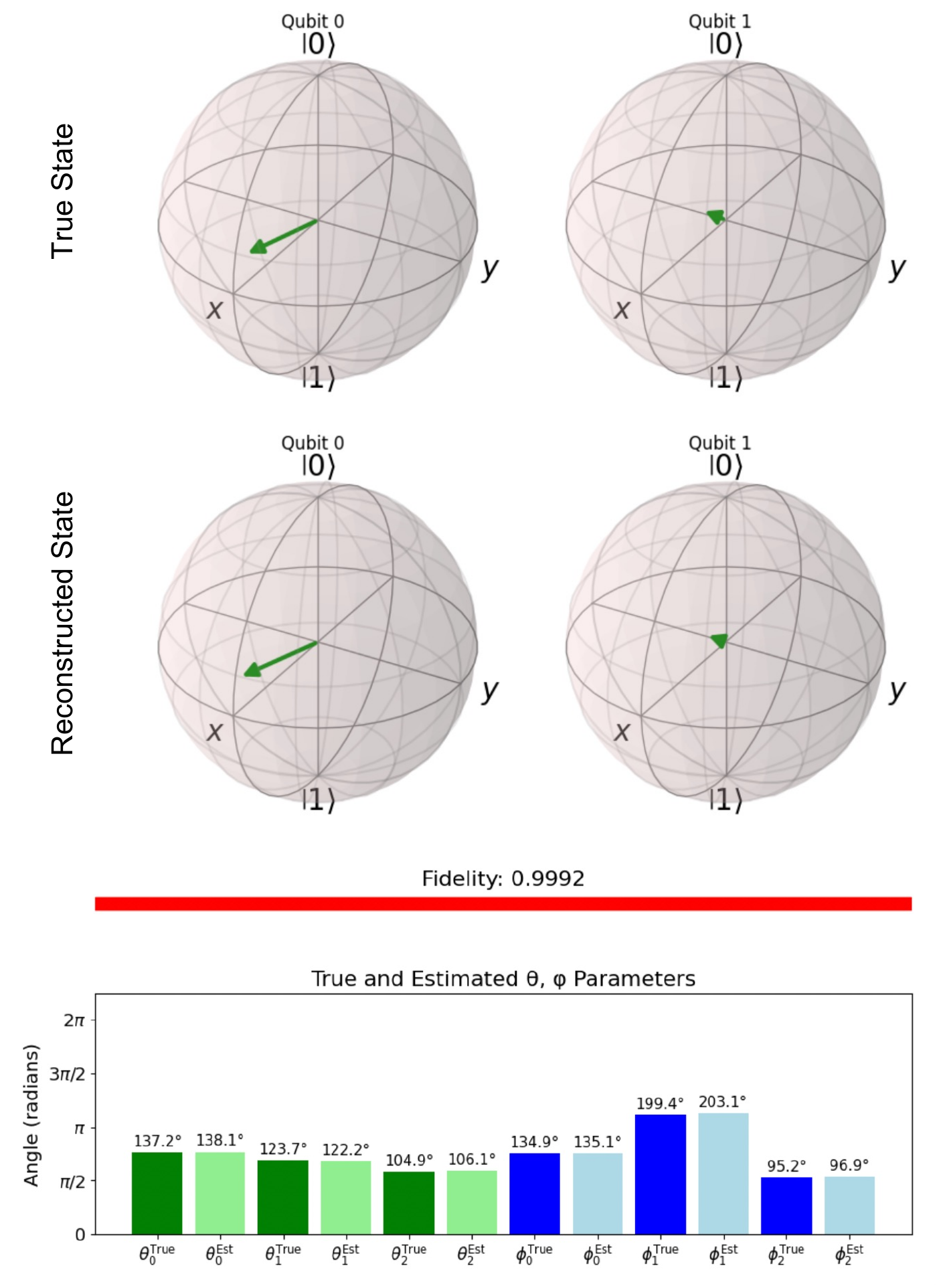}
    \caption{Visualization of two-qubit state tomography using QubitLens. The Bloch spheres show the true (top) and reconstructed (bottom) reduced states of each qubit. The bar plot compares the true and estimated values of the polar parameters. In this example, measurements are performed in all three Pauli bases 
$\bfX,\bfY,\eqand\bfZ$ for each qubit. }
    \label{fig:QubitLensMultiQubit}
\end{figure}

QubitLens automatically performs the necessary partial traces and visualizes both the true and reconstructed reduced states of each qubit using Qiskit's \texttt{qiskit\_bloch\_multivector}\cite{qiskit_bloch_multivector} function. In addition to the geometric visualization, the interface displays the estimated polar parameters \((\hat{\theta}_j, \hat{\phi}_j)\) and compares them against the true parameters through a side-by-side bar plot. The fidelity between the reconstructed and true global states is also reported as a horizontal gauge bar.
Fig.~\ref{fig:QubitLensMultiQubit} shows an example visualization for a 2-qubit system in QubitLens, where the true and reconstructed Bloch vectors are displayed for each qubit along with the global fidelity and the parameter-wise comparison.

\noindent \emph{Code Availability.}
The full implementation of the {QubitLens}, including both the single-qubit and multi-qubit tomography modules, is available as open-source code in the GitHub repository: \url{https://github.com/mdaamirQ/QubitLens}.

\bibliographystyle{IEEEtran}
\bibliography{references}

\end{document}